\documentclass[12pt]{article}
=0cm \headsep =0cm

\newcommand{\be}{\begin{equation}}
\newcommand{\ee}{\end{equation}}
\newcommand{\bea}{\begin{eqnarray}}
\newcommand{\eea}{\end{eqnarray}}

\usepackage {latexsym}
\usepackage {graphicx}

\begin{document}

\begin{titlepage}
\title{Bulk photons in Asymmetrically Warped Spacetimes and
Non-trivial Vacuum Refractive Index}

\author{K. Farakos$^b$ \footnote{kfarakos@central.ntua.gr}, N. E. Mavromatos$^a$
\footnote{nikolaos.mavromatos@kcl.ac.uk} and P. Pasipoularides$^b$ \footnote{paul@central.ntua.gr} \\
      $^a$King's College London, University of London\\
       Department of Physics, Strand WC2R 2LS, London, U.K.
       \\
       $^b$Department of Physics, National Technical University of
       Athens \\ Zografou Campus, 157 80 Athens, Greece}
\date{ }
       \maketitle

\begin{abstract}
We consider asymmetrically warped brane models, or equivalently
brane models where the background metric is characterized by different time and
space warp factors. The main feature of these models is that 4D Lorentz symmetry
is violated for fields which propagate in the bulk, such as gravitons. In this
paper we examine the case of bulk photons in asymmetrically warped brane models.
Although our results are general, we examine here two specific but characteristic solutions:
1) AdS-Schwarzschild 5D  Black Hole solution and 2) AdS-Reissner Nordstrom 5D Black
Hole solution. We show that the standard Lorentz invariant dispersion relation
for 4D photons is corrected by nonlinear terms which lead to an Energy-dependent speed
of light. Specifically, we obtain a sub-luminous Energy-dependent
refractive index of the form $n_{eff}(\omega)=1+c_{G}\;\omega^2$, where $\omega$ is the energy of
the photon, and the factor $c_G$ is always positive and depends on the free parameters
of the model. Finally, comparing the results with recent data from the MAGIC Telescope,
claiming a delayed arrival of photons from the Active Galactic Nucleus of Mk501,
we impose concrete restrictions to the two sets of models examined in this work.
We shall also discuss briefly other possible astrophysical constraints on our models.
\end{abstract}
\end{titlepage}
\tableofcontents

\section{Introduction and Motivation}

Theorists, in an attempt to solve the hierarchy problem, invented new string theory
models with relatively large extra  dimensions. The early realization~\cite{antoniadis}
that the string scale is an arbitrary parameter, and can be as low as a TeV scale, lead naturally to the consideration of models with large extra dimensions ~\cite{ArkaniHamed:1998rs,Antoniadis:1998ig}, introducing the so-called \emph{brane-world models}, in which the standard model particles are assumed to be localized in a 3D
brane (our world), embedded in a multi-dimensional manifold (bulk).
Subsequently, models in which the bulk space time is warped have been proposed~\cite{Randall:1999ee,Randall:1999vf}.
In such models, the extra dimensions could be: (i)  finite, if a second parallel brane world lies at a finite bulk distance from our world~\cite{Randall:1999ee}, thus providing a new hierarchy of mass scales,
or: (ii) infinite, if our world is viewed as an isolated brane, embedded in an (infinite) bulk space~\cite{Randall:1999vf}. In fact, it is the presence of such warp factors that provides~\cite{Randall:1999ee} in the case (i) a ``resolution'' to the hierarchy problem. The important problem of stabilization of the extra dimension, in the case of two brane models, has been considered in Refs. \cite{radion}.

Such brane or string models lead to no apparent contradictions with the present-day observations regarding four dimensional physics \cite{Rubakov:2001kp}, although many falsifiable predictions can be made of relevance to either
particle physics colliders, such as LHC and future linear colliders, or astrophysical/cosmological measurements.
In the standard brane world scenario, only gravitons are allowed to propagate in the bulk, contrary to
the standard model particles, which are constrained to lie on the brane. We call this the \emph{Standard String/Brane-World} scenario (SSBW).

Beyond the SSBW scenario, there are models
in which all or some of the standard model particles,
can live in the bulk. The original scenario of universal extra dimensions, where standard model particles possess KK excitations, was presented
for compactified dimensions, $R\leq 10^{-17} cm$, in~\cite{dienes}; the connection with brane worlds of Randall-Sundrum (RS) type~\cite{Randall:1999ee} can be achieved
by considering as ``bulk'' the available space between the two parallel branes. Note, that in the case of the first RS model the size of the extra dimension is very short: in particular it is of the order of the Planck length.
In such a case, gauge fields and fermions are not necessarily localized on
the brane, see for example \cite{Davoudiasl:2007wf} and references therein.
For our purposes in this article, we call this second scenario the \emph{Extended String/Brane-World} scenario (ESBW).

There are numerous generalizations of the above generic models, including, for instance, topological defects
along the extra dimension(s), or higher-order curvature corrections, see for example Refs. \cite{PerezLorenzana:2005iv, Maartens:2003tw, Mavromatos:2000az,Mavromatos:2002vt, Giovannini:2001ta, Mavromatos:2005yh, BarbosaCendejas:2007hs, Saridakis:2007wx, Bertolami:2007dt, Andrianov:2007tf,veneziano}. Also, there are models in which the standard model particles are localized on the brane dynamically, via a mechanism which is known as \emph{layer-phase mechanism}, see for example Refs.
\cite{Farakos:2005hz,Farakos:2006tt,Farakos:2006sr,Farakos:2007ua,Pasipoularides:2007zz,Dimopoulos:2006qz} and references there in. Moreover, an effective propagation of standard model particles in the bulk may characterize the so-called ``fuzzy'' or fluctuating-thick-brane-world scenarios~\cite{campbell}, according to which our brane world
quantum fluctuates in the bulk. In such a case, there are uncertainties in the bulk position of the brane world,
resulting in an ``effectively'' thick brane~\cite{lizzi}. In such a situation, even the standard SSBW
scenario would result  in an ``effective'' propagation of the standard model particles in the bulk. In such ``effectively thick'' brane models~\cite{lizzi}, the wavefunction of a standard model particle has an extent  in the bulk region, the latter being defined as the ``effective''
thickness of the ``fat brane'',  of string (or Planck) size.

In the previous part we presented a short review on brane world models, in what follows we focus to the so called
asymmetrically warped brane models, and we give the essential references for the understanding of this paper.
In particular, we will consider the following generic ansatz for the metric in five dimensions
\be
ds^2=-\alpha^{2}(z)dt^2+\beta^{2}(z)d\textbf{x}^2+\gamma^2(z) dz^2
\label{asymm}
\ee
where $z$ parameterizes the extra dimension, $\alpha(z)$ is the time warp factor and $\beta(z)$ is
the three-dimensional-space warp factor~\footnote{Although one can set $\gamma$ to one by
a coordinate transformation in static models, this is not the case in more general cases with a time dependent
$\gamma (z,t)$, which are of interest in cosmology~\cite{Chung:2000ji}, and which we make use here.
It is in this sense we base our
analysis here in the frame where the 5D metric has the form (\ref{asymm}).}. Models which are described
by metrics of the form of Eq. (\ref{asymm}), in which the time and three-dimensional-space warp factors are different, are often called asymmetrically warped brane models. In these models, although the induced metric on the brane (localized at $z=0$) is Lorentz invariant
upon considering  the case $\alpha(0)=\beta(0)$,  the metric of Eq. (\ref{asymm}) does not
preserve 4D Lorentz invariance in the bulk since $\alpha(z)\neq\beta(z)$ for $z\neq 0$.

Models with equal warp factors, such as the RS model~\cite{Randall:1999ee,Randall:1999vf},
have so far attracted the main attention, since 4D Lorentz invariance is assumed
as a fundamental symmetry of nature. However, brane models with
asymmetrically warped solutions, like that of Eq. (\ref{asymm}) above,
have also been constructed~\cite{Chung:2000ji, Visser:1985qm, Csaki:2000dm, Dubovsky:2001fj, Bowcock:2000cq}.
We note here that the static \textit{asymmetric} solutions that are presented
in Refs.~\cite{Visser:1985qm, Csaki:2000dm, Dubovsky:2001fj, Bowcock:2000cq}
presuppose the existence of extra matter in the bulk~\footnote{For example, the solution or
Ref.~\cite{Visser:1985qm} requires a constant electric field along the extra dimension, and a fine tuning
similar to that of the RS model.},
hence these models luck the simplicity of construction characterising the RS-model.

Although symmetric ~\footnote{With the terminology \textit{symmetric }
solutions we mean here metrics with equal space and time warp factors,
or equivalently metrics which possess SO(3,1) symmetry, like that of
RS model.} brane models are the most important, as they incorporate exact 4D Lorentz symmetry, at least in our opinion there is an argument in favor of asymmetrically warped
space-times, presented in detail in Ref. \cite{Chung:2000ji}.
Static solutions, like that of RS model, can be used approximately
only for a short period of time around our cosmological epoch $t_0$.
For larger periods the complete cosmological evolution must be considered.
In such a cosmological context the metric depends on the comoving time coordinate t, and has the form:
\be
ds^2=-\alpha^{2}(z,t)dt^2+\beta^{2}(z,t)d\textbf{x}^2+\gamma^2(z,t)dz^2 \label{assym1}
\ee
Note, that the above time depended solutions have in general different space and time warp factors.
The corresponding static metric, which describes our universe in the short
time period humans exist, which is very small compared to the cosmological time scale of the
evolution of the Universe, is obtained if we set $t=t_0$ in Eq. (\ref{assym1}).
According to this philosophy, in general, the metric of Eq. (\ref{assym1}) evolves to an
asymmetrically warped solution for $t=t_0$, or equivalently we expect that  $\alpha(z,t_0)\neq\beta(z,t_0)$.
Although, we have given arguments on the possibility that such vacua can indeed characterize cosmological brane models, nevertheless we cannot say with certainty that they cannot decay to standard Minkowski vacua. However, since they cannot be ruled out on theoretical grounds, at least presently, it worths imposing constraints on their parameters phenomenologically, and this is what we aim to do in this article.

The question then arises as to how one can constrain or exclude/falsify brane models with \textit{asymmetric}
solutions of the form of Eq. (\ref{asymm}), on account of
present (or immediate-future) experimental bounds on (local) Lorentz symmetry violation in the bulk.
In the SSBW case, where the bulk is completely unaccessible by
the standard model particles, Lorentz violation signals can
be observed only by bulk particles in the gravitational sector,
like gravitons, or at most
particles neutral under the standard model group, e.g. right-handed sterile neutrinos.
Bulk fields can "see"
the asymmetry between the warp factors in the extra dimension,
whilst standard-model particles, which are rigidly ``pinned'' on the brane,
can only "see" equal warp factors $\alpha(0)=\beta(0)$.
Gravity effects which could reveal 4D Lorentz violation are described
in Ref. \cite{Csaki:2000dm} (see also Refs. \cite{Cline:2001yt, Cline:2002fc, Cline:2003xy}) where superluminous propagation of gravitons is
possible for specific models with \textit{asymmetric} solutions.
However, since the detection of gravitons is still not an experimental fact, such Lorentz violations
are still compatible with the current experiments, both terrestrial and astrophysical, probably awaiting the
future detection of gravitational waves in order to be constrained significantly.

However, in the ESBW scenario, where some or all of the
standard model particles, are allowed to propagate in the bulk,
such Lorentz-Invariance-violating effects can be bounded by high precision tests
of Lorentz symmetry, since now, 4D Lorentz violation may
be revealed even in the standard model sector. In this way, stringent restrictions
to \textit{asymmetric} models can be imposed by astrophysical
observations and other high-energy experimental tests.

In this paper, we aim to study the propagation of bulk photons in an
\textit{asymmetrically} warped metric background by
solving the (classical) equations of motion for a 5D massless
U(1) Gauge field in the curved background of Eq. (\ref{asymm}).
We shall demonstrate that the standard Lorentz invariant dispersion relation
for 4D photons possesses nonlinear corrections, which lead to an
energy-dependent speed of light on the brane. Specifically,
we shall obtain a sub-luminal refractive index for photons
$n_{eff}(\omega)=1+c_G\;\omega^2$, where $\omega$ is the energy of
the photon, and the factor $c_G$ is always positive and depends on the free parameters
of the models.
Finally, comparing these results with the data of the Magic experiment on a delayed arrival
of highly energetic photons from the distant galaxy Mk501~\cite{magic,NM} we shall set concrete
restrictions to our models. We shall also discuss briefly other astrophysical constraints on our models.

\section{5D U(1) Gauge fields in asymmetrically warped spacetimes}

In this section we study bulk photons in an asymmetrically warped metric
background of the form (\ref{asymm}). Although, our conclusions are valid
for the quite general ansatz of Eq. (\ref{asymm}), nevertheless it is convenient to demonstrate our results
for specific classes of solutions to the 5D Einstein Equations. In what follows, we shall
consider a slightly modified RS-metric
which is asymmetrically warped, and it satisfies the bulk Einstein equations. We shall consider a class of solutions
which are known as 5D AdS-Reissner-Nordstrom Black Hole Solutions~\cite{Csaki:2000dm}. These solutions are presented in detail in the
following subsection, 2.1. In subsection 2.2 we show that is possible to write the 5D AdS-Reissner-Nordstrom Solution
as a linearized perturbation around the RS metric. Then, it is not difficult to mimic the $S^{1}/Z_2$ structure
of the first RS-model~\cite{Randall:1999ee}. However, the novel feature of our model is that the background
metric deviates slightly from the standard RS-metric, and as a result there is a small violation
of 4D lorentz symmetry in the bulk. In section
2.3 we study the propagation of bulk photons in the above-mentioned metric background.
At this stage, we consider it as instructive to remind the reader that the extra dimension in the first RS-model
is assumed to be compact and very small, of the order of the Planck length ($\ell_P \sim 10^{-33}cm$). Hence,
our assumption of bulk Gauge field is not in conflict with
the current observations in particle accelerators, as we have also emphasized in the introduction.

\subsection{5D AdS-Reissner-Nordstrom Solutions }

We consider an action  which includes 5D gravity, a negative cosmological
constant $\Lambda$, plus a bulk U(1) Gauge field~\cite{Csaki:2000dm}:
\begin{equation}
S=\int d^5 x \sqrt{g}\left(\frac{1}{16 \pi G_5}(R^{(5)}-2\Lambda)-
\frac{1}{4}B^{MN}B_{MN} \right)+\int d^4 x \sqrt{g_{(brane)}}{\cal L}_{matter}, \label{matter} \end{equation}
where $G_{5}$ is the five dimensional Newton constant, and $B_{MN}=\partial_{M}H_N-\partial_{N}H_{M}$
is the field strength of the U(1) Gauge field $H_M$, with $M,N = 0,1 \dots 5$.
Note, that this additional bulk Gauge field does not interact with
charged matter on the brane, so it must not be confused with the usual electromagnetic field $A_M$, representing a bulk photon,
which will be introduced in the next subsection. The four dimensional term
in the action corresponds to matter fields localized on the
brane, which is assumed located at $r=r_0$, and are described by a \emph{perfect fluid}
with energy density $\rho$  and pressure $p$. This brane term is necessary for the
solution of Eq. (\ref{csaki0}) to satisfy the junction conditions on the brane, as we will explain in detail in the next section.
The corresponding Einstein equations can be written as\footnote{In the case of a second brane at the position $r'_0$ we need an additional  $\delta(r-r'_0)$ term in equation (\ref{einstein1}).}
\begin{equation}
G_{MN}+\Lambda g_{MN}=8\pi G_5\left(\delta(r-r_0)\frac{\sqrt{|g^{(brane)}|}}{\sqrt{|g|}} T^{(matter)}_{\mu\nu}
\delta_{M}^{\mu}\delta_{N}^{\nu}+T_{MN}^{(B)} \right)\label{einstein1}
\end{equation}
where the energy momentum tensor for the U(1) Gauge field is:
\begin{equation}
T_{MN}^{(B)}=B_{MP}B_{N}^{~P}-\frac{1}{4}g_{MN} B_{PS}B^{PS}
\end{equation}
For the metric of the black hole solution we make the ansatz
\begin{equation}
ds^2=- h(r) dt^2+\ell^{-2}r^2d\Sigma^2+h(r)^{-1}dr^2  \label{csaki0}
\end{equation}
where $d\Sigma^2=d\sigma^{2}+\sigma^{2}d\Omega^2$ is the metric of the spatial 3-sections, which
in our case are assumed to have zero curvature, in agreement with the current astrophysical phenomenology, pointing
towards spatial flatness of the observable Universe. Moreover, $\ell$ is the AdS radius which is equal to
$\sqrt{-\frac{6}{\Lambda}}$. By solving Einstein equations (\ref{einstein1}) we obtain:
\begin{equation}
h(r)=\frac{r^2}{\ell^2}-\frac{\mu}{r^2}+\frac{Q^2}{r^4}
\end{equation}
where $\mu$ is the mass (in units of the five dimensional Planck scale) and $Q$ the charge of the 5D \emph{AdS-Reissner-Nordstrom} black hole.
This, of course, presupposes the
existence of extra bulk matter, namely a point-like source with mass $\mu$ and charge $Q$.
Note that, in the case of nonzero charge $Q$, a non-vanishing component $B_{0r}$ of the bulk field-strength tensor $B_{MN}$:
\begin{equation}
B_{0r}=\frac{\sqrt{6}}{\sqrt{8\pi G_5}}\frac{Q}{r^3}~,
\end{equation}
is necessary
in order for the solution to satisfy the pertinent Einstein-Maxwell equations.

In the limit $Q=0$, which corresponds to the 5D \emph{AdS-Schwarzschild} Black Hole, the bulk vector field $H_M$ becomes redundant.
In this article we shall also examine this case, in connection with effects on photon propagation in 4D and, as we shall see,
the presence of the extra charge in the Reissner-Nordstrom case does not affect qualitatively the conclusions regarding the order of magnitude of the asymmetry between the warp factors in order to
get consistency with the current phenomenology.

\subsection{Construction of two-brane models and Null Energy Condition}

We will describe now how one can use the 5D AdS-Reissner-Nordstrom vacuum in order to construct brane models. As a first step, we place a brane at the position $r=r_{0}$. We next assume that for $r>r_0$ the 5D metric is given by Eq.~(\ref{csaki0}), while for $r<r_0$ the metric is given by Eq.~(\ref{csaki0}) upon the replacement $r\leftrightarrow r_0^2/r$. The metric obtained this way  is $Z_2$-symmetric upon the replacement $r\leftrightarrow r_0^2/r$. The next step is to glue the two independent slices of the metric by including a perfect fluid energy momentum tensor of the form
\begin{equation}
T_{\mu}^{\;\nu}={\rm Diag}(-\rho,p,p,p)
\end{equation}
and satisfying the Israel junction conditions on the brane, where $\rho$ is the energy density and $p$ is the pressure. The equation
of state is parametrised as usual by:  $\omega=p/\rho$.

We remind the reader, that 5D black
holes suffer from a singularity at $r=0$, which may be naked or not, depending on the magnitudes of the parameters $Q,\mu,\ell$. In particular, if $Q^4<\frac{4}{27}\mu^3\ell^2$ the 5D black hole has two horizons \cite{Csaki:2000dm}.
In this case, a single brane construction is possible as a horizon can isolate the brane from the singularity. Contrary to the Randall-Sundrum (RS)-model, this
class of brane models has the advantage that one does not need a fine tuning between the energy density $\rho$ on the brane and the 5D cosmological constant. This self-tuning property of the AdS-Reissner-Nordstrom black holes serves as
a possible solution to the cosmological constant problem \cite{Csaki:2000dm}.

However, if we demand the existence of a horizon, the matter we have to put on the brane violates the null energy condition
$\rho+p\geq 0$  (or $\omega\geq -1$), see Ref \cite{Csaki:2000dm}. This is a quite embarrassing situation, as we have to introduce unconventional sources like ghosts. In Refs. \cite{Cline:2001yt,Cline:2002fc} the authors present a complete investigation on this issue including cases with: a) the presence of external fields in the bulk, b) the existence of
higher-derivative terms in the action of Gauss-Bonnet type and c) models with the $Z_2$ symmetry relaxed.  The result of such analyses is a \emph{no-go theorem}, according to which it is impossible to shield the singularity from the brane by a horizon, unless the null energy condition $\rho+p\geq 0$ is violated.

Nevertheless, in the current work we are not concerned with a solution to the cosmological constant problem. Our purpose is simply to construct brane models which incorporate a Lorentz symmetry violation due to the difference of the space and time warp factors. In this case we can isolate the singularity from the brane by putting a second brane at a position $r=r_0'$ (between the first brane and the singularity). Hence, we have a two-brane construction similar to the first RS-model, in which the two branes are located at the fixed points of an orbifold $S_1/Z_2$. Note, that because the Israel junction conditions must be satisfied also on the second brane, we have to assume an additional energy momentum tensor of a perfect fluid, with energy density $\rho'$, pressure $p'$ and $\omega'=p'/\rho'$, localized on the second brane. We also consider the physically interesting case in which $\epsilon\equiv r_0'/r_0\sim 10^{-6}$ where
the large hierarchy between the Planck and electroweak scale is satisfied. However, in the case of two-brane models fine-tuning relations between 4D energy densities and 5D dimensional cosmological constant are unavoidable.

The question that arises at this point is wether two-brane constructions
that respect the null energy condition on the branes do exist. An analysis towards this direction can be found in Ref. \cite{Cline:2003xy}. We have also performed independently a similar analysis, which confirms these results, but it is quite lengthy to be presented in this paper, as this would take one out of the main point of this work. For our purposes here, it suffices to state that construction of several classes of two-brane models in the 5D AdS-Reissner-Nordstrom vacuum is possible, without violation of the null energy condition, and hence two-brane constructions are acceptable.

\subsection{5D AdS-Reissner-Nordstrom Solution as a linearized perturbation around the Randall-Sundrum metric }

In order to achieve our aim, which is to write the 5D AdS-Reissner-Nordstrom
solution as a \emph{linearized perturbation} around the RS metric, we perform
the following change of variables $r \to z(r)$ in Eq. (\ref{csaki0}):
\begin{eqnarray}
r&=& r_0 e^{-k\;z}~, \quad {\rm for}~  z>0 \nonumber \\
r&=& r_0 e^{k\;z}~, \quad  {\rm for}~ z<0~,
\end{eqnarray}
If, in addition,
we make the rescaling $x_{\mu}\rightarrow \frac{r_0}{\ell}x_\mu \quad (\mu=0,\dots 3)$,  we obtain:
\bea
ds^2=-a^2(z) h(z) dt^2+a^2(z)d\textbf{x}^2+h(z)^{-1}dz^2  \label{csaki}
\eea
where $a(z)=e^{-k|z|}$, and $k=\ell^{-1}$ is the inverse $AdS_5$ radius. For the function $h(z)$ we obtain:
\begin{equation}
h(z)=1-\delta h(z), \quad \delta h(z)=\frac{\mu \ell^2}{r_0^4} e^{4 k|z| }-\frac{Q^2 \ell^2}{r_0^6}e^{6k|z|} \label{deviation}
\end{equation}
As we describe in detail in section 2.2, it is not difficult to construct two brane models.
Now, the $Z_2$ symmetry $r\leftrightarrow r_0^2/r$ if it expressed in the frame of the new parameter $z$, reads $z\rightarrow -z$. In addition, the positions of the branes which are located at $r_0$ and $r'_0=r_0 \epsilon$ in the original coordinate system, in the new coordinate system are determined by the equations $z=0$ and $z=\pi r_c$ correspondingly, where $\epsilon=e^{-k\pi r_c}$ and $r_c$ is radius of the compact extra dimension.

Finally, we assume that $|\delta h(z)|\ll 1$ in the interval $0<z<\pi r_c$, or equivalently we adopt that $\delta h(z)$ is
only a small perturbation around the RS-metric. This implies that $r_0$, which is the radius that determines the position
of the brane in the bulk, is (comparatively) a very large quantity. In particular, we have to satisfy both the following two
inequalities $ r_0^2 \epsilon^{2}  \gg \sqrt{\mu} \ell$ and $r_0^3 \epsilon^3 \gg Q \ell$.

\subsection{Equation of motion for the Bulk Photon and Non-Standard Vacuum Refractive Index}

In this subsection we will study the case of a 5D massless $U(1)$ gauge boson $A_{N}$ in the background of an
asymmetrically warped solution of the form of Eq. (\ref{csaki}). We stress that the gauge field $A_{N}$ must not be
confused with the gauge field $H_{N}$, introduced in the previous section.
As we will see later, we will identify the zero mode of $A_{N}$ with
the standard four dimensional photon. On the other hand $H_{N}$ is an additional bulk field which
does not interact with the charged particles on the brane. The corresponding equation of motion for $A_{N}$ reads:
\be
\frac{1}{\sqrt{g}}\partial_{M}\left(\sqrt{g} g^{MN}g^{RS}F_{NS}\right)=0~, \label{photon}
\ee
with $F_{NS}=\partial_{N}A_S-\partial_{S}A_M$, and $N,S= 0,1,\dots 5$.
In the background metric of Eq. (\ref{csaki}), Eq.~(\ref{photon}) gives:
\bea
-\partial_z(a^{2}(z) h(z) \partial_z A_j)-\nabla^{2}A_j+\frac{1}{h(z)}\partial_{0}^{2}A_j=0, \quad  j=1,2,3~,
\eea
where we have considered the Coulomb gauge condition:
\be
\vec{\nabla }\cdot \vec{A}=0, \quad A_{0}=0,\quad A_{z}=0 ~, \label{coulomp}
\ee
which is suitable for the case of a Lorentz violating background.
On setting in Eq. (\ref{photon}):
\be
A_j(x,z)=e^{i p \cdot x}\chi_j(z), \quad p_{\mu}=(-\omega,\textbf{p})
\ee
and keeping only the linear terms in $\delta h(z)$, we obtain
\be
-\partial_{z}\left\{a^{2}(z) \left[1-\delta h(z)\right]\partial_z\chi\right\} +\left\{\textbf{p}^2-\left[1+\delta h(z)\right]\omega^2\right\}\chi=0 \label{sch1}
\ee
where for brevity we have dropped the index $j$ from $\chi$.
Note that the \emph{spectrum} of Eq.~(\ref{sch1}) is \emph{discrete}, due to the orbifold boundary conditions~\cite{Randall:1999ee},
$\chi'(0)=0$ and $\chi'(\pi r_c)=0$ (where the prime denotes a $z$-derivative). Hence we add from now on an index $n$ $(n=0,1,2,3,..)$,
denoting the $n_{th}$ eigenstate
of the equation.

Upon applying the formalism of time-independent perturbation theory, Eq. (\ref{sch1}) can be written as
\be
(H_{0}+\Delta H) \chi_n=m_n^2 \; \chi_n, \quad m^2=\omega^2-\textbf{p}^2 \label{schl2}~.
\ee
The hermitian operator
\be
H_0=-\partial_{z}\left(a^{2}(z) \partial_z\right)~,
\ee
which is of zeroth order in $\delta h(z)$, corresponds to the unperturbed Hamiltonian, while the hermitian operator
\be
\Delta H=-\delta h(z) \: \omega^2+\partial_z(a^{2}(z) \delta h(z) \partial_z ) \label{schl23}
\ee
which is linear in $\delta h(z)$, corresponds to the perturbation Hamiltonian.

The quantity $\Delta H$ is indeed a small perturbation when compared to $H_0$, since the relative discrepancy $\delta h(z)$
between the space and time warp factors is assumed to be very small ($|\delta h(z)|<<1$) in the interval $[0,\pi r_c]$.

We need to find a complete set of eigenfunctions $\chi_{n}^{(0)}$ ($n=0,1,2,...$) of the unperturbed Schr\"odinger equation
\be
H_{0} \chi_{n}^{(0)}=\left(m^{(0)}_{n}\right)^2 \chi_{n}^{(0)} \label{unper}
\ee
It can be seen straightforwardly that for $n=0$,
\begin{equation}
m_{0}^{(0)}=0~,
\label{zeromode}
\end{equation}
is
an eigenvalue of Eq. (\ref{unper}), or equivalently
Eq. (\ref{unper}) possesses a \emph{zero mode}, which corresponds to the 4D \emph{photon}.
The corresponding wave function is constant in the interval $[0,\pi r_c]$.
For our purposes here we shall omit the detailed derivation of the eigenvalue spectrum of (\ref{unper}), as this
has been done previously in Refs.~\cite{Davoudiasl:1999tf, Pomarol:1999ad}, where we refer the
interested reader for details. It suffices to mention that the nonzero eigenvalues are:
\begin{equation}
m^{(0)}_{n}=x_{n}\:k\:e^{-k\pi r_{c}}, \: n=1,2,3,... \label{eigen}
\end{equation}
where $x_{n}$ are the roots of the zeroth order Bessel function $J_{0}(x_n)=0$. On adopting $k r_{c}\sim 12$, which is
the standard choice in order to connect electroweak (ew) and Planck scales ($M_P=e^{k\pi r_c} m_{ew} $) in a RS
framework~\cite{Randall:1999ee},
one obtains that:
\begin{equation}
m^{(0)}_{n}\sim {\rm TeV}~, \qquad n = 1,2,\dots ~.
\label{mtev}
\end{equation}
The corresponding eigenfunctions are:
\begin{eqnarray}
&&\chi_{0}^{(0)}=\frac{1}{N_0}, \quad \quad \quad \quad \quad \quad N_0=\frac{1}{\sqrt{\pi r_c}}  \label{eigen1}\\
&&\chi_{n}^{(0)}=\frac{1}{N_n} e^{k z}J_{1}(\frac{m^{(0)}_n}{k}e^{kz}), \quad N_n=\frac{e^{k\pi r_c}}{\sqrt{2 k}}J_1(x_n) \label{eigen2}
\end{eqnarray}
where the coefficients $N_0, \; N_n$ are defined by the normalization condition:
\begin{equation}
\int_{0}^{\pi r_c} \chi_{n}^{(0)}(z)\chi_{m}^{(0)}(z) dz=\delta_{mn}
\end{equation}
The lowest energy eigenvalue $m_{0}^2$ of Eq. (\ref{schl2}) can be expanded as:
\be
 m_{0}^2=\left(m^{(0)}_{0}\right)^2+\left(m^{(1)}_{0}\right)^2+\left(m^{(2)}_{0}\right)^2+... \label{exp}
\ee
where the zeroth order correction (\emph{zero mode}) is $m_0^{(0)}=0$ (c.f. (\ref{zeromode}).
The first and second order corrections are obtained by the formulae of time-independent perturbation theory:
\be
(m_{0}^{(1)})^2=\langle \chi_{0}^{(0)}|\Delta H|\chi_{0}^{(0)}\rangle \label{first}
\ee
\be
(m_{0}^{(2)})^2=\sum_{n\neq 0}\frac{|\langle \chi_{0}^{(0)}|\Delta H|\chi_{n}^{(0)}\rangle|^2}{\left(m^{(0)}_0\right)^2-\left(m^{(0)}_n\right)^2} \label{second}
\ee
Upon using Eqs. (\ref{exp}), (\ref{first}) and (\ref{second}), we obtain:
\be
\omega^2-\textbf{p}^2=-a_G\;\omega^2-b_G \;\omega^4 \label{dispr0}
\ee
with
\be
a_G=\int_{0}^{\pi r_c}dz\left(\chi_{0}^{(0)}(z)\right)^2 \delta h(z) \label{bg0}
\ee
\be
b_G=\sum_{n\neq 0}\frac{1}{\left(m_{n}^{(0)}\right)^2}\left(\int_{0}^{\pi r_c}dz\: \chi_{0}^{(0)}(z)\chi_{n}^{(0)}(z)\delta h(z)\right)^2 \label{bg}
\ee
Note, that the derivative term in (\ref{schl23}) does not contribute to $a_G$ and $b_G$, as it acts on a constant eigenfunction (zero mode).
The parameters $a_G$ and $b_G$ are independent from the (photon) energy $\omega$, and $b_G$ is always positive. Eq. (\ref{dispr0}) then gives:
\be
|\textbf{p}|=\omega\sqrt{(1+a_{G})+b_G \omega^2} \label{dis}
\ee
According to Eq. (\ref{dis}), the phase velocity of light is equal to
\be
v_{ph}=\frac{\omega}{|\textbf{p}|}=\frac{1}{\sqrt{(1+a_{G})+b_G \omega^2}}
\ee
or equivalently, for $b_G \omega^2<<1$:
\be
v_{ph}\simeq \frac{1}{\sqrt{(1+a_{G})}}-\frac{b_G}{2(1+a_{G})^{\frac{3}{2}}}\omega^{2} \label{dispr2}
\ee
where $\omega$ is the energy of the photon.

On the other hand, for the photon's group velocity we have:
\be
v_{gr}=\frac{d\omega}{d|\textbf{p}|}=\frac{\sqrt{(1+a_{G})+b_G \omega^2}}{(1+a_{G})+2 b_G \omega^2}
\ee
and the corresponding formula for $b_G<<1$ is
\be
v_{gr}\simeq \frac{1}{\sqrt{(1+a_{G})}}-\frac{3 \; b_G}{2(1+a_{G})^{\frac{3}{2}}}\omega^{2} \label{dispr3}
\ee
One can define the constant velocity of light \emph{in standard vacuo} as the low energy limit ($\omega \to 0$) of the phase velocity of Eq. (\ref{dispr2})
\be
c_{light}= \frac{1}{\sqrt{(1+a_{G})}} \label{dispr4}
\ee
This definition is reasonable, if we take into account that the velocity of light is measured in practice only
for low energy photons. We need very high energies, in our case of order  $10^3$ TeV or higher, in order for the contribution of
the energy dependent term in Eq. (\ref{dispr2}) to be significant, exceeding the accuracy with which the velocity of light is presently known.

From Eqs. (\ref{dispr2}) and (\ref{dispr4}) we then obtain an
\emph{effective subluminal refractive index} for the \emph{non-standard vacuum} provided by our brane-world constructions:
\begin{equation}
n_{eff}(\omega)=\frac{c_{light}}{v_{ph}}=1+\frac{b_G}{2(1+a_G)}\omega^2 ~.\label{effective}
\end{equation}
This is the main result of this paper. An issue, we would like to emphasize, is that the phase and group velocities
(Eqs. (\ref{dispr2}) and (\ref{dispr3}) respectively), as well as  the effective refractive index
of Eq. (\ref{effective}), depend quadratically on the photon energy $\omega$. Moreover, we note that equation (\ref{effective}) is a
perturbative result which is valid only for energies $\omega<<1/\sqrt{b_G}$. For larger energies (for example in the case of ultra high energies cosmic rays $\omega\sim 10^{20}~{\rm eV} $), Eq. (\ref{effective}) is not valid and a full nonperturbative treatment is necessary. In this limit the subluminal nature of the refractive index may be lost.

According to the refractive index (\ref{effective}),
photons with different energies propagate with different velocities.
Due to the sub-luminal nature of the refractive index there will be delays in the
arrival times of more energetic photons, as compared with the lower-energy ones,
provided of course that simultaneous emission of the various modes is assumed at the source.
Hence, we will observe a time lag of the arrival times of photons
which were emitted simultaneously by a remote astrophysical source,
with the lower energy photons arriving first.  It is worth noticing that the
effective refractive index (\ref{effective}) is independent on the polarization of photons, or
equivalently our model does not predict birefringence. If we had included higher order corrections to the electromagnetic
lagrangian (e.g. in the form of a Born Infeld lagrangian) birefringence may occur. In the next section (section 3) we compare our result with the data
of Magic experiment in order to set bounds to our model.

\section{Setting bounds: Comparison with MAGIC and other Data}

In the previous section we examined the case of bulk photons in asymmetrically warped brane models and we derived an equation  (Eq. (\ref{effective})) which implies a nontrivial refractive index for 4D photons.
In this section we aim to use this modified quadratic dispersion relation in order to set bounds to our model, via a direct comparison with the experimental data of Magic. We have used the MAGIC Telescope observation of photons from Mk501 \cite{magic,NM}, because this is the only case at present of an observed effect on a delayed photon arrival, with the higher energy photons arriving later. The constraints we derive, in this section, for the parameters of our model, are the most stringent on quadratically modified  photon dispersion relations, as far as direct photon observations are concerned. Note, that there are other experiments which set bounds on quadratically modified  photon dispersion relations, however they do not give more stringent constraints than Magic effect. For example in subsection 3.2, we compare with the recent data from H.E.S.S. Telescope \cite{hess}, and as we see the magic constraints are the most stringent. Finally, in subsection 3.2, we discuss briefly possible constraints to our model which are not based on the quadratic dispersion relation of Eq. (\ref{effective}), including the cases of the fermions and the ultra high energy cosmic rays.                                                                                                                                                    

\subsection{Comparison with MAGIC Data}

We wish now to compare the
time delays of the more energetic photons, implied by the dispersion relation (\ref{effective}),
with experimentally available
data. In view of the quadratic suppression of the non trivial vacuum effects with the
Planck scale, it becomes evident that the higher the energy range of the photons detected, the higher the
sensitivity of the pertinent experiment to such effects.
Recently, experimental data from observations of
the MAGIC Telescope~\cite{magic} on photon energies in the TeV range have become available.
It therefore makes sense to compare the time delays predicted in our models, due to (\ref{effective}),
with such data, thereby imposing concrete restrictions
on the free parameters of the models.

For completeness let us first review briefly the
relevant observations~\cite{magic,NM}. MAGIC
is an imaging atmospheric Cherenkov telescope which
can detect very high energy (0.1 TeV-30 TeV) electromagnetic
particles, in particular gamma rays. Photons with very
high energy (VHE) are produced from conversion of
gravitational energy at astrophysical distances from Earth, when, say, a very massive rotating star is
collapsing to form a supermassive black hole. Astrophysical objects
like this are called blazars and are active galactic nuclei (AGN).

The observations of MAGIC during a flare (which lasted twenty minutes)
of the relatively nearby (red-shift $z \sim 0.03$) blazar of Markarian (Mk) 501 on July 9 (2005), indicated a
$4\pm 1$ min time delay between the peaks of the time profile envelops
of photons with energies smaller than 0.25 TeV and photons
with energies larger than 1.2 TeV. Possible interpretations
of such delays of the more energetic photons have already been proposed.
Conventional (astro)physics at the source may be responsible for the delayed emission of
the more energetic photons, as a result, for instance, of some non-trivial versions of the
Synchrotron Self Compton (SSC) mechanism~\cite{magic}. It should be noted at this stage that the
standard SSC mechanism, usually believed responsible for the production of VHE photons in other AGN, such as Crab Nebula, fails~\cite{magic} to explain the results of MAGIC, as a relative inefficient acceleration is needed in order
to explain the ${\cal O}({\rm min})$ time delay.
Modified SSC models have been proposed in this respect~\cite{magic2}, but the situation is not conclusive.
This prompted speculations that new fundamental physics, most likely quantum-gravity effects,
during propagation of photons from the source till observation,  may be responsible for inducing the observed delays, as a result of an effective refractive index for the vacuum, see for example
Ref.~\cite{Ellis:1999sd} and references therein.

Exploring the fact that MAGIC had the ability of observing individual photons, a numerical analysis on the relevant
experimental data has been performed in~\cite{NM}, which aimed at the reconstruction
of the original electromagnetic pulse by maximizing its energy upon the assumption of a sub-luminal vacuum refractive index
with either linear or quadratic quantum-gravity-scale suppression:
\begin{equation}
n_{eff}(\omega)=1+\left(\frac{\omega}{M_{QG(n)}}\right) ^{n}, \; \quad n=1,2 \label{Magic}
\end{equation}
The analysis in \cite{NM} resulted in the following values for the quantum-gravity mass scale at 95 \% C.L.
\begin{equation}
M_{QG(1)}\simeq 0.21 \;~ 10^{18} ~{\rm GeV}, \quad M_{QG(2)}\simeq 0.26 \;~ 10^{11} ~{\rm GeV} \label{Magic0}
\end{equation}
We must emphasize at this point that many Quantum Gravity Models seem to predict
modified dispersion relations for probes induced by vacuum refractive index effects, which appear
to be different for each quantum gravity approach, not only
as far as the order of suppression by the quantum gravity scale is concerned,
but also its super- or sub-luminal nature. Some models, for instance, entail birefringence effects,
which can be severely constrained by astrophysical measurements~\cite{crab}.
There are also alternative models, see for example Refs. \cite{Gogberashvili:2006mr,Carmona:2002iv}.

In \cite{NM1}, a non-perturbative mechanism for the observed time delays has been proposed, based on
stringy uncertainty principles within the framework of a
string/brane theory model of space-time foam. The model entails a brane world crossing regions in bulk space time
punctured by point-like D-brane defects (D-particles). As the brane world moves in the bulk,
populations of D-particles cross the brane, interact with photons, which are attached on the brane world (SSBW scenario),
and thus  affect their propagation. The important feature of the stringy uncertainty delay mechanism
is that the induced delays are proportional to a single power of the photon energy $\omega$, thus being
linearly suppressed by the string mass scale, playing the quantum gravity scale in this model.
It is important to notice that the mechanism of \cite{NM1} does not entail any modification of the
local dispersion relations of photons. In fact the induced delays are also independent of photon polarization, so
there are no gravitational birefringence effects. It is also important to notice that, in view of
the electric charge conservation, the D-particle-foam defects can interact non trivially only with photons or at most electrically neutral
particles and no charged ones, such as
electrons, to which the foam looks transparent.

As we have showed in the previous section, a
sub-luminal vacuum refractive index may characterize
asymmetrically warped brane-world  models, assuming propagation of photons in the bulk.
Our model, however, predicts a refractive index with quadratic dependence
on energy, c.f. (\ref{effective}).
Comparing Eqs. (\ref{effective}) with (\ref{Magic}), we obtain:
\begin{equation}
\frac{b_{G}}{2(1+a_{G})}\leq M_{QG(2)}^{-2} \label{Magic1}
\end{equation}

The parameters $a_G$ and $b_{G}$ can be computed numerically by means of Eqs. (\ref{bg0}) and (\ref{bg}). Note,
that the deviation $\delta h(z)$, the eigenvalues $m_{n}^{(0)}$ and the eigenfunctions $\chi_{n}^{(0)}$
are known analytically, see Eqs. (\ref{deviation}), (\ref{eigen}), (\ref{eigen1}) and (\ref{eigen2}) above.
We have computed numerically the parameter $b_{G}$ for two exemplary cases: a) for an AdS-Reissner-Nordstrom Solution (\ref{deviation}),
and b) for an AdS-Schwarzschild Solution (obtained by setting $Q=0$ in Eq. (\ref{deviation}))  .
In the former case we assume for simplicity that $\mu$ and $Q^2/r_{0}^2$ are of the same order of magnitude (for details see Ref. \cite{Cline:2003xy} or the discussion in section 2.2, see also Eq. (\ref{fine}) below),
and hence we can ignore the contribution of the $e^{4 k z}$ term in Eq. (\ref{deviation}).
In particular we obtain:
\begin{eqnarray} &&b_{G}\simeq 2.95\: \langle \delta h  \rangle^{2}\:~ {\rm TeV}^{-2}, \quad {\rm Ads-Reissner-Nordstrom}  \label{Magic3} \\
&&b_{G}\simeq 10 \: \langle \delta h  \rangle^{2}\: ~ {\rm TeV}^{-2}, \quad \:\:\: {\rm Ads-Schwarzschild}  \label{Magic2}~,
\end{eqnarray}
where  $\langle \delta h  \rangle$ is defined as the average value:
\begin{equation}
\langle \delta h  \rangle=\frac{1 }{\pi r_c}\int_{0}^{\pi r_c}\delta h(z) dz~.
\end{equation}
We will use  $\langle \delta h  \rangle$ in order to estimate the degree of violation of Lorentz symmetry in our models.
Taking into account Eqs. (\ref{Magic0}), (\ref{Magic1}) ,(\ref{Magic3}) and (\ref{Magic2}) we find the constraints:
\begin{eqnarray}
&&\mid\langle \delta h  \rangle\mid \leq 1.4 \;10^{-8} \;\quad {\rm Ads-Reissner-Nordstrom}  \label{Magic5} \\
&&\mid\langle \delta h  \rangle\mid \leq 0.75 \;10^{-8} \quad {\rm Ads-Schwarzschild}  \label{Magic4}~.
\end{eqnarray}
The small values we obtain are consistent with the weak nature of $\langle \delta h  \rangle$, as required
by treating it as a perturbation. Also we observe that the values for the average deviation of Eqs. (\ref{Magic3}) and (\ref{Magic5})
are of the same order of magnitude for both Ads-Schwarzschild and Ads-Reissner-Nordstrom solutions.
In the above analysis we ignored the effects due to the Universe expansion, since the latter do not affect the order of magnitude
of the above bounds due to the small red shift ($z \sim 0.03$) of Mk501 we restrict our discussion in this section.
The inclusion of such effects, which are essential for larger redshifts,
is straightforward~\cite{NM} and does not present
any conceptual difficulty.

Note, that although the parameter $a_G$ is not important for our analysis, it is crucial when we have to make comparisons with the velocities of other particles.
If we take into account Eq. (\ref{bg0})  we see that $a_G=\langle \delta h\rangle $. Hence, the parameter $a_G$ is constrained via the equation
\begin{equation}
\mid a_G \mid\leq10^{-8}
\end{equation}
This summarizes the constraints to the asymmetrically-warped brane models with bulk photons using the data of the MAGIC experiment.

\subsection{Comparison with other data}

 However it is worth to be mentioned that another Telescope, H.E.S.S., looking at a flare from another celestial object, the Active Galaxy PKS 2155-304
at a much larger distance (red-shift $z = 0.116$), and emitting photons in the few-TeV energy range
with a much higher (unprecedented) statistics, has not claimed an observed delayed arrival of the most energetic photons. The associated limit
in the quadratic refractive index from such observations was~\cite{hess}:
\begin{equation}\label{hess2}
     |\zeta | < 7.3 \times  10^{19} ~, \qquad n_{eff}(\omega) = \left(1 + \zeta \frac{\omega^2}{M^2_{\rm P}}\right)
\end{equation}
with $M_{\rm P} = 1.22 \times 10^{19} $ GeV, the conventional Planck mass scale in the notation of Ref.~\cite{hess}. Using (\ref{effective}) we can translate these bounds to bounds for the parameters of our model, in an entirely analogous way with the MAGIC case (\ref{Magic5},\ref{Magic4}) studied above.
Because of the larger distance from Earth, however, of the AGN PKS 2155-304 as compared to Mk501, the effects of the expansion of the Universe~\cite{Ellis:1999sd,NM} must be taken into account in this case, when fitting the results to the observed arrival times of photons. We shall not do this in this work, since the bounds obtained from
the H.E.S.S. measurements (\ref{hess2}) are not better from the ones obtained from MAGIC, discussed above. Indeed,  in the notation (\ref{hess2}) the lower bound of the quantum gravity scale from these null measurements (as far as delayed arrivals are concerned), assuming quadratic modifications in the photon dispersion relations, is~\cite{hess}
\begin{equation}
M_{\rm QG(2)} >  M_{\rm P}/|\zeta|^{1/2} \sim 1.4 \times 10^{9}~{\rm GeV}~,
\end{equation}
which is worse than the corresponding MAGIC bound (\ref{Magic0}), and hence for our purposes here the MAGIC bound is the most stringent one based on direct high-energy astrophysical photon time-of-flight measurements.

However, in the models examined in this paper the modified dispersion relations are expected to characterize also other massive particles that may be allowed to propagate in the bulk.
One may in this way impose more severe restrictions, for instance, by comparing the velocity of bulk photons with that of other massive particles, such as massive fermions. For such a comparison, additional considerations are necessary: a) the fermions are localized in one of the two branes or b) the fermions can propagate in the bulk,  a case which has not been analyzed in this paper. If, for example, one finds that fermions move faster than light, then severe additional restrictions have to be imposed, due to the emission of vacuum Cherenkov radiation, for details see Ref. \cite{crab}. However an investigation toward this direction is a topic of a completely new work.

Moreover, in case a photon exhibits modified dispersion relations, as advocated here, then scattering of Highly Energetic photons (with energies higher than $10^{19}$ eV) of soft photons of the Cosmic Microwave Background ($\gamma_{\rm HE} + \gamma_{\rm CMB} \to e^+ e^-$) may lead to entirely different physics than in the Lorentz-invariant case, as this affects the energy thresholds for pair production. This may lead to severe constraints on the relevant models~\cite{sigl}. Naively, one could use the
bounds, which are presented in Ref. ~\cite{sigl}, to set limits in the parameters of our model.
However, the analysis of such very high energy photons requires a nonperturbative analysis going beyond the approximations made in the present paper, and is left for future investigation.

\section{Comparison with the velocity of gravitons}

Finally, before closing we would like to compare our results for bulk photons with the velocity of gravitons, which are also allowed to propagate in the bulk.  In Refs. \cite{Csaki:2000dm, Cline:2001yt} the following relation is found:
\begin{equation}
c_{graviton}^{(bulk)}=1+\frac{Q^2 \ell^2}{4 r_0^6} (\epsilon^{-4}-2 \epsilon^{-2})
\end{equation}
where we have adopted the position of Ref. \cite{Cline:2003xy}, in the case where the null energy condition on the brane is not violated,  for $w=-1$: \begin{equation}
\mu=\frac{3}{2}\frac{Q^2}{r_0^2} \label{fine}
\end{equation}
On the other hand, as discussed here, bulk photons propagate with a velocity given by Eq. (\ref{dispr4}), or equivalently:
\begin{equation}
c_{light}^{(bulk)}=1-\frac{\langle \delta h\rangle}{2}, \quad \langle \delta h\rangle=a_G
\end{equation}
where
\begin{equation}
\langle \delta h\rangle=\frac{1}{2 k \pi r_c} \left(\frac{\mu \ell^2}{2 r_0^4}(\epsilon^{-4}-1)-\frac{Q^2 \ell^2}{3 r_0^6}(\epsilon^{-6}-1)\right), \quad \epsilon=e^{- k \pi r_c}\sim 10^{-16} \label{dh}
\end{equation}
Taking into account Eq. (\ref{fine}) we find
\begin{equation}
c_{light}^{(bulk)}=1+\frac{Q^2 \ell^2}{12 k \pi r_cr_0^6} \left(\epsilon^{-6}-\frac{3}{2}\epsilon^{-4}\right)~.
\end{equation}
Then, the difference reads:
\begin{equation}
\delta c_{(lg)}= c_{light}^{(bulk)}-c_{graviton}^{(bulk)}\simeq \frac{Q^2 \ell^2}{12 k \pi r_cr_0^6} \epsilon^{-6} \quad (>0)
\end{equation}
We observe that if photons propagate in the bulk then gravitons are not superluminous. However, we can not use this result to set constraints
via gravitational Cherenkov radiation, since the velocity of bulk photons is in general different from the velocities of fermions, which we have not computed in this work. For more details on this topic the interested reader is referred to Refs. \cite{ Cline:2003xy, Moore:2001bv}.

\section{Conclusions}

In this article we have examined non-standard Lorentz-violating vacua, induced in the brane-world framework due to
asymmetrically warped bulk space times. We have considered the effects of such an asymmetric warping on four-dimensional photons, with wavefunctions that extend into the bulk. These assumptions lead to an effective refractive index of the 4D
non-standard vacuum arising in this context, which scales quadratically with the quantum gravity (length) scale and photon
energy. By comparing our results with data from the MAGIC experiment, pointing towards a delayed arrival time of more
energetic photons from Mk501, as compared to lower-energy ones, we have derived bounds on the size of the
perturbations around the classical solutions, or order $10^{-8}$ (c.f. (\ref{Magic5},\ref{Magic4})). This
result seems rather generic, independent of the particular type of the assumed bulk solution.

In order to set bounds, we have used the MAGIC Telescope observation of photons from Mk501, because this is the only case at present of an observed effect on a delayed photon arrival, with the higher energy photons arriving later. The constraints we derive, in section 3, are the most stringent on quadratically modified  photon dispersion relations, as far as direct photon observations are concerned. We would like to mention that there are also other experiments which set bounds on quadratically modified  photon dispersion relations. In particular, in subsection 3.2, we compare with the recent data from H.E.S.S. Telescope \cite{hess}, and we see that the magic constraints are the most stringent. Finally, in subsection 3.2, we discuss briefly possible constraints to our model which are not based on the quadratic dispersion relation of Eq. (\ref{effective}), including the cases of fermions and ultra high energy cosmic rays. However, an analysis towards these directions is quite extended to be included in this work, and is left for further investigation.

\section*{Acknowledgements}

The work of N.E.M. is partially supported by the European Union through the FP6 Marie-Curie Research and Training Network \emph{Universenet}
(MRTN-2006-035863).

\end{document}